\documentclass[12pt]{iopart}

\usepackage{graphicx}
\usepackage{dcolumn}
\usepackage{bm}
\usepackage{gensymb} 
\usepackage{siunitx}
\usepackage[utf8]{inputenc}
\usepackage[T1]{fontenc}
\usepackage{mathptmx}
\usepackage{array, multirow, bigdelim, makecell, booktabs} 
\newcommand{\ch}[1]{\ensuremath{\,\mathrm{#1}}}

\begin{document}

\title[Growth of AlN on a SiN substrate for Hybrid Photonic Circuits]{Growth of Aluminum Nitride on a Silicon Nitride Substrate for Hybrid Photonic Circuits}
\author{G Terrasanta$^1$ $^2$, M M{\"u}ller$^3$ $^1$, T Sommer $^1$ $^4$, S Gepr\"{a}gs$^3$, R Gross$^3$ $^1$ $^4$, M Althammer$^3$ $^1$ and M Poot$^1$ $^4$ $^5$}

\address{$^1$ Department of Physics, Technical University Munich, Garching, Germany}
\address{$^2$ Physics Section, Swiss Federal Institute of Technology in Lausanne (EPFL), Switzerland}
\address{$^3$ Walther-Mei\ss ner-Institut, Bayerische Akademie der Wissenschaften, 85748 Garching, Germany}
\address{$^4$ Munich Center for Quantum Science and Technology (MCQST), Munich, Germany}
\address{$^5$ Institute for Advanced Study, Technical University Munich, Garching, Germany}
\eads{\mailto{menno.poot@tum.de}}


\begin{abstract}
Aluminum nitride (AlN) is an emerging material for integrated quantum photonics with its excellent linear and nonlinear optical properties. In particular, its second-order nonlinear susceptibility $\chi^{(2)}$ allows single-photon generation. We have grown AlN thin films on silicon nitride (\ch{Si_3N_4}) via reactive DC magnetron sputtering. The thin films have been characterized using X-ray diffraction, optical reflectometry, atomic force microscopy, and scanning electron microscopy. The crystalline properties of the thin films have been improved by optimizing the nitrogen to argon ratio and the magnetron DC power of the deposition process. X-ray diffraction measurements confirm the fabrication of high-quality c-axis oriented thin films with a full width at half maximum of the rocking curves of 3.9$\degree$ for 300-nm-thick films. Atomic force microscopy measurements reveal a root mean square surface roughness below \SI{1}{nm}. The AlN deposition on SiN allows us to fabricate hybrid photonic circuits with a new approach that avoids the challenging patterning of AlN.
\end{abstract}
\noindent{\it Reactive sputtering, aluminum nitride, silicon nitride, photonic integrated circuits, hybrid integration, hybrid photonic circuits}

\section{\label{sec:introduction}Introduction}
AlN thin films are widely employed in the fabrication of electromechanical devices. Their large piezoelectric coefficient, resistivity, and their compatibility with silicon process technology \cite{trolier2004thin} make them excellent candidates for many different applications, such as micro-electro-mechanical systems (MEMS) resonators \cite{Piazza2006} and microwave filters \cite{Dubois1999}. Recently, AlN has been employed as a promising material for integrated photonics \cite{wan2020large,xiong2012aluminum}: its wide bandgap and strong second-order nonlinear susceptibility $\chi^{(2)}$ allowed for the realization of nonlinear processes in low-loss and high-quality resonators, from the ultra-violet to infrared wavelengths\cite{pernice2012,guo2017parametric,Liu_OpEx_AlN_on_sapphire_UV,Jung:14}. Achieving highly c-axis oriented films is essential for obtaining high-quality nonlinear optical properties \cite{Larciprete2006}. AlN films have been fabricated with several different techniques, including molecular beam epitaxy \cite{Dasgupta2009}, chemical vapor deposition \cite{chiang2011}, and pulsed laser deposition \cite{szekeres2011}. Here, we show the optimized growth of AlN on high-stress SiN via reactive DC magnetron sputtering. This technique has the advantages of being low-cost as well as low-temperature \cite{Iqbal2018}. Our interest in depositing c-axis thin films on SiN stems from the use of this material as the main platform for the fabrication of our photonic circuits \cite{hoch2020chip,poot2016design,poot2016characterization}. By sputtering AlN on top of pre-patterned SiN substrates, we fabricate novel hybrid AlN/SiN photonic circuits \cite{Terrasanta2021}. The thickness, density and crystallographic orientation of the sputtered films are evaluated using X-ray diffraction (XRD). These properties are first optimized for growth on \ch{SiO_2} on (001)-oriented Si substrates, before growing AlN on high-stress SiN. Afterwards, the growth results on SiN are presented and compared to those on \ch{SiO_2}. Moreover, the refractive index is obtained via spectral reflectance, while the root mean square (RMS) surface roughness is evaluated with atomic force microscopy (AFM) and scanning electron microscopy (SEM). For photonic applications, the AlN films need to have a refractive index similar to SiN and their surface roughness needs to be as low as possible to reduce light losses due to Rayleigh scattering \cite{xiong2012aluminum}. Both properties were successfully achieved, allowing for the fabrication of low-loss hybrid photonic circuitry. Finally, the composite AlN/SiN fabrication is explained and hybrid structures are characterized.

\section{\label{sec:exp}Sample fabrication}
\begin{figure}[ht]
\centering
\includegraphics[width=0.87\textwidth]{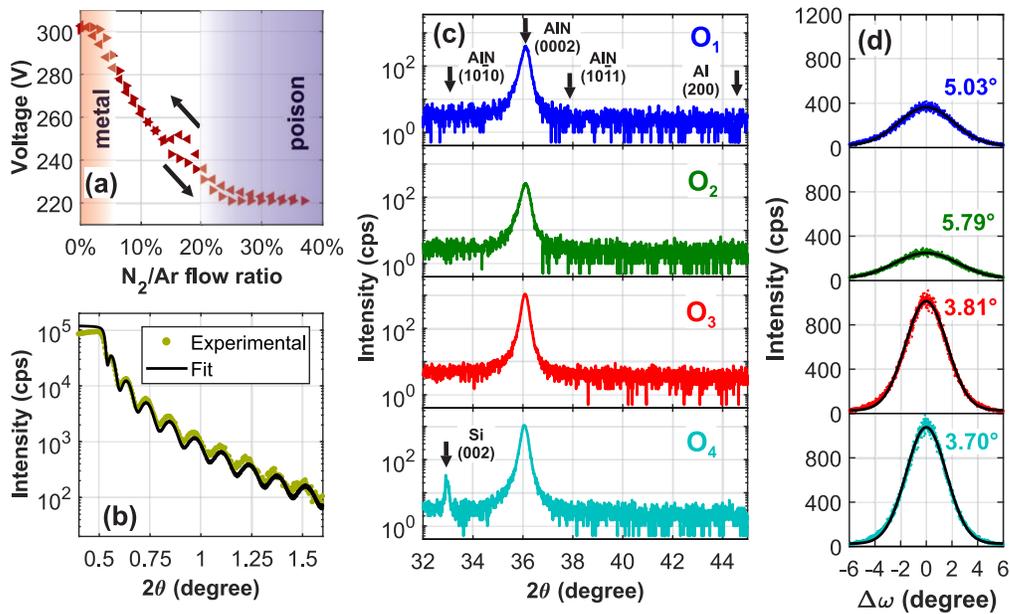}
\caption{\label{fig:sputterSiO2} (a) Characterization of the AlN reaction states (metal, transition, poison) by plotting the magnetron voltage versus the \ch{N_2}/Ar flow ratio during the sputtering deposition. The metal regime (red shaded area) is at a very low ratio, then the transition takes place and the poison regime (blue shaded area) is reached at about 30\%. (b) Low-angle X-ray reflectometry of an AlN thin film grown on \ch{SiO_2}. The experimental data are fitted using the \textit{Leptos} simulation software with film thickness and density as parameters, resulting in \SI{59}{nm} and \SI{3.23}{g/cm^3}, respectively. 
$2\theta$-$\omega$ X-ray diffraction scans (c) and rocking curves (d) of samples $O_1$, $O_2$, $O_3$, and $O_4$ with AlN grown on \ch{SiO_2}. The $2\theta$ positions of the AlN~($10\bar{1}0$), the AlN~(0002), the AlN~($10\bar{1}1$), as well as the Al~(200) reflections are indicated by vertical arrows. In sample $O_4$ a faint Si~($002$) signal coming from the substrate is visible at $33\degree$. Sputtering conditions and X-ray diffraction results of the samples are summarized in Table \ref{tab:table1deposition}. The rocking curves are shifted to \SI{0}{\degree} for comparison. cps: counts per second.}
\end{figure}

AlN thin films are grown via DC reactive magnetron sputtering of a 3-inch Al target in a mixed nitrogen (\ch{N_2}) and argon (Ar) atmosphere. The base pressure of the sputtering chamber before deposition is lower than \SI{5e-9}{mbar}. The sample is inserted and heated to a deposition temperature of \SI{600}{\celsius}. Once the desired temperature is reached, the reactive gases \ch{N_2} and Ar are introduced into the chamber and the deposition pressure is set to \SI{5e-3}{mbar}. The Al target is pre-sputtered for 15 minutes to provide a reproducible starting state. During pre-sputtering the sample is protected by a shutter. The sample is positioned face-down at a distance of \SI{4}{cm} from the target and the shutter is kept open for the estimated deposition time in order to reach the desired thickness. The deposited AlN is formed by a chemical reaction between \ch{N_2} and the sputtered Al. This reaction takes place on exposed surfaces of the chamber, as well as the target and the sample, while reactions in the gas phase are unlikely \cite{depla2019}. The AlN reaction is characterized by two different regimes: a metal regime in which only a few Al atoms react with \ch{N_2} and mostly purely metallic Al is deposited, and a poison regime, where the Al reacts with \ch{N_2} and AlN is deposited on the sample surface as wanted \cite{Iqbal2018}. It is important to determine the transition between these regimes to select the optimal \ch{N_2}/Ar flow ratio. The transition from the metallic to poison state is measured with the following procedure: the power and deposition pressure are set to \SI{70}{\watt} and \SI{5e-3}{mbar}, respectively. The total flow of Ar plus \ch{N_2} is set to a constant value of \SI{20}{sccm}, while the \ch{N_2}/Ar flow ratio is varied from 0 to 37\% with both increasing and decreasing \ch{N_2}-flow steps of 0.2 sccm. At each step, the magnetron DC voltage is recorded after 30 seconds of waiting time. As shown in Figure \ref{fig:sputterSiO2}(a), the voltage has a plateau at low \ch{N_2} to \ch{Ar} flow ratio, then decreases for increasing \ch{N_2} flow, and finally saturates at around 30\%. As explained by Iqbal et al.\cite{Iqbal2016}, the plasma impedance decreases when AlN is formed at the cathode surface due to the higher emission rate of secondary electrons. Therefore, the voltage drops when Al and \ch{N_2} start to bind together and it saturates once the Al target is completely covered with AlN \cite{Iqbal2016}. This measurement, thus, enables us to identify the \ch{N_2}/Ar ratio for the metal-, transition- and poison-state of the magnetron. In the first two regimes, \ch{Ar^{+}} would dominate in the vicinity of the target, resulting in an Al-AlN mixed-phase, but with a larger kinetic energy transfer to the Al atoms. In contrast, nitrogen atoms dominate at higher concentration and pure AlN is deposited, but the transferred energy is reduced due to the lower mass of nitrogen with respect to argon \cite{Iqbal2016}. The crystalline structure of the deposited AlN films may, thus, be different at the onset of, and deep in the poison regime.

\begin{table}
\lineup
\footnotesize
\caption{\label{tab:table1deposition}Summary of sputtering conditions and characterization results of different AlN thin films. All films are fabricated at \SI{600}{\celsius} and \SI{5e-3}{mbar} in a custom-made sputtering system from \textit{Bestec GmbH}. Two different substrates are employed: Samples with $O_i$ IDs are grown on substrates with \SI{1500}{nm} wet \ch{SiO_2} on (001)-oriented Si from \textit{Micro Chemicals}, whereas $N_i$ samples are deposited on \SI{330}{nm} high-stress LPCVD \ch{Si_3N_4} on \SI{3300}{nm} \ch{SiO_2} on (001)-oriented Si from \textit{Active Business Company}. The bracket indicates samples that are sputtered in the same run. Thicknesses and refractive indices are obtained via spectral reflectance, and RMS roughnesses using AFM. Rates are calculated by dividing the obtained thicknesses by the deposition times. The (0002) peak intensities and FWHM are obtained from the $2\theta$-$\omega$ scan and the rocking curve, respectively.}
\begin{tabular}{@{}ccccccccccccc} 
\br
ID & Ar & N$_2$ & Power & Time & Thickn. & Rate  & (0002)  & (0002)   & Refractive & RMS \\
 & flow & flow & (W) & (s)  & (nm) & (nm/min) & intensity & FWHM  & index &  roughn. \\ 
& (sccm) & (sccm) & &   & & &(cps) & ($ \degree $) & @\SI{632.8}{nm} & (nm) \\ \mr
$O_1$  & 10 & 2.5 & \070 & 1815 & 190& \06.3 & \0406 & 5.03 & 2.05 & 0.30 \\ 
$O_2$  & 10 & 5.0 & \070 & 2288 & 185& \04.9 & \0267 & 5.79 & 2.05 & 0.40 \\ 
$O_3$  & 10 & 2.5 & 210 & \0562 &  248 &26.5& 1093 & 3.81 & 2.08 & 0.79 \\ 
$O_4$  & 10 & 3.5 & 210 & \0784 & 238& 18.2 & 1091 & 3.70 & 2.09 & 0.84\\  
\hspace{-1em}\rdelim\{{2}{*}[] $O_5$  & 10 & 3.5 & 210 & \0986 &  285 & 17.3 & 1186 & 3.70 & 2.03 & 0.64 \\
$N_1$  & 10 & 3.5 & 210  & \0986 &297 & 18.1& 1088 & 3.93 & 2.00 & 0.47\\ 
$N_2$  & 10 & 3.5 & 210 & \0704 & 204& 17.3 & \0455 & 4.93 & 2.01 & 0.26\\ 
$N_3$  & 10 & 3.5 & 210 & \0388 & 110& 16.9 & \0174 & 5.90 & 2.04 & 0.23 \\ 
$N_4$  & 10 & 3.5 & 210 & \0265 & \074 & 16.7& \0149 & 6.81 & 2.03 & 0.26\\ \br 
\end{tabular}
\end{table}

\section{Results}
\subsection{X-ray diffraction}
The crystal structure of the films is characterized using X-ray diffraction. The measurements are carried out on a four-circle diffractometer equipped with a standard Cu X-ray source using a Ni-filter to suppress Cu $K_\beta$-radiation. First, the thickness and density of the sputtered thin films are evaluated using low-angle X-ray reflectometry. Figure \ref{fig:sputterSiO2}(b) shows an example of such a measurement together with a simulation to fit the experimental data. This measurement is useful to determine the film density and verify the sputtering rate at the start of the optimization process. The averaged density is 3.22$\pm$\SI{0.06}{g~cm^{-3}}, which is in agreement with the bulk value of \SI{3.23}{g~cm^{-3}}~ \cite{levinshtein2001properties} within the margin of error. 
Next, $2\theta$-$\omega$ X-ray diffraction scans are performed to evaluate the crystal structure under different sputtering conditions. As an example, the scans of samples $O_1$, $O_2$, $O_3$, and $O_4$ are shown in Figure \ref{fig:sputterSiO2}(c). These samples are part of the optimization series for the deposition of AlN on SiO2, and their sputtering conditions are summarized in Table \ref{tab:table1deposition}.
A priory, it is not clear if AlN or Al will be deposited, nor what the orientation of the former will be. The four peaks that may therefore manifest within the scanned interval are the AlN~($10\bar{1}0$) reflection at \SI{33.25}{\degree}, the AlN~(0002) reflection (c-axis orientation) at \SI{36.09}{\degree}, the AlN~($10\bar{1}1$) reflection at \SI{37.97}{\degree}, and the Al~(200) reflection at \SI{44.76}{\degree} \cite{Downs2003}. Here the standard crystallographic four-index notation of hexagonal crystal structures is used. We note, however, that in literature on AlN growth the pseudo-cubic notation with three Miller indices labeling the planes is also commonly used. There, the AlN reflections are denoted as (100), (002), and (101), respectively. 
As apparent from Figure 1(c), each sample displays a clear AlN~(0002) reflection, which demonstrate c-axis-oriented growth of the AlN thin films, without any signals from other AlN reflections or from unwanted Al. Note, that in sample $O_4$ another, much smaller, reflection peak is visible at \SI{32.95}{\degree}, which is close to the expected position of AlN~($10\bar{1}0$) at \SI{33.25}{\degree}. A closer inspection shows, however, a sharp profile with full width at half maximum (FWHM) of \SI{0.14}{\degree}. This is a factor of two narrower than for the AlN~(0002) peaks. Hence, this peak is identified as the Si~(002) reflection expected at \SI{32.99}{\degree}. Although forbidden, this peak can be visible when the Si substrate experiences finite strain or bending.

In the optimization process, the influence of the \ch{N_2}/Ar ratio is first tested by increasing the \ch{N_2} flow from sample $O_1$ to $O_2$, bringing the process well into the poison regime [Figure~\ref{fig:sputterSiO2}(a)]. This, however, results in a decrease of the intensity of the AlN~(0002) reflection. A possible reason comes from the fact that each of the AlN crystal orientations requires different energies to form, where the (0001) orientation requires the highest energy \cite{Xu2001}. The higher concentration of nitrogen results in less energy transfer to the sputtered atoms \cite{Iqbal2016}, thus deteriorating the (0001) orientation. Nevertheless, the \ch{N_2} flow still needs to be high enough to allow for the poison state of the magnetron. A second optimization approach is to increase the magnetron DC voltage while maintaining the same gas flow ratio ratio as in $O_2$. This is done by increasing the DC magnetron power for sample $O_3$, which shows a significantly higher AlN (0002) peak. The increased voltage, thus, improves the c-axis orientation of the AlN thin film. Finally, the nitrogen flow is increased for the fabrication process of sample $O_4$. The increased flow does not affect the peak intensity as before, because the increased DC voltage now gives a stronger contribution to the energy transfer.
This conjecture is confirmed by rocking curve measurements. The FWHM of a rocking curve is a measure of the quality of the crystalline structure \cite{xrd}. The FWHMs of the AlN~(0002) reflections shown in Figure~\ref{fig:sputterSiO2}(d) are obtained by applying a Gaussian fit with a constant offset that accounts for the background intensity. The results of the fits are indicated in the plot and listed in Table \ref{tab:table1deposition}. Sample $O_4$ has the smallest FWHM of all AlN thin films deposited on SiO2 with 3.70$\degree$ at \SI{238}{nm} AlN thickness, thus confirming the improved degree of crystal orientation. This is comparable with literature results. For instance, other research groups have sputtered thick AlN films, above \SI{1.5}{\mathbf{\mu}m}, with a FWHM as low as 1.6$\degree$, which resulted in a second-order nonlinear coefficient of $d_{33} = \SI{11}{pm/V}$ \cite{Larciprete2006}. Also thinner films grown on molybdenum layers have shown a high degree of crystal orientation, with a FWHM of the AlN~(0002) reflection below 2$\degree$  for a film thickness of \SI{300}{nm} \cite{Felmetsger2011}. In another work, second-order nonlinear optical processes were realized in AlN thin films with a FWHM of 2$\degree$ of the AlN~(0002) reflection in \SI{330}{nm} thick optical waveguides\cite{xiong2012aluminum}. \\

\begin{figure}[ht]
\centering
\includegraphics[width=0.88\textwidth]{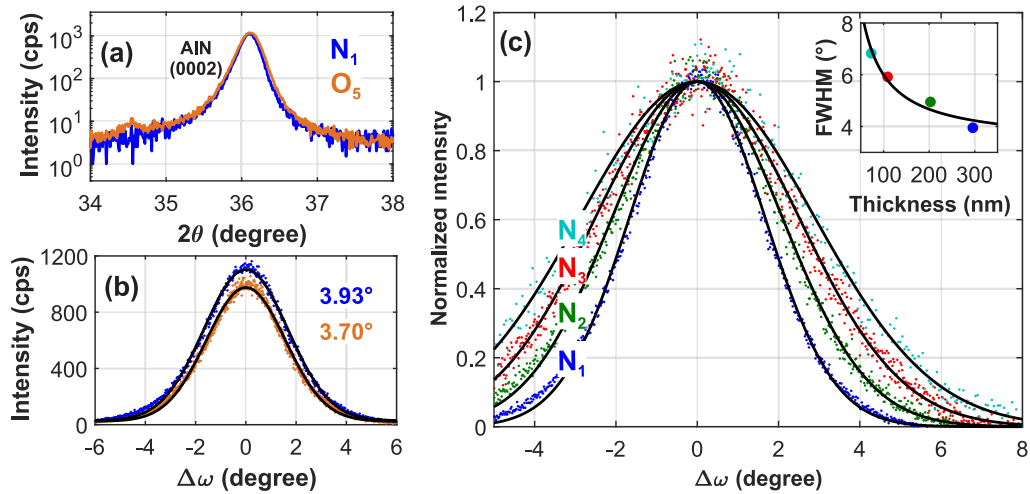}
\caption{\label{fig:compare_SiN_and_SiO2} Comparison of the $2\theta$-$\omega$ scan (a) and the AlN~(0002)-reflection rocking curve (b) between AlN sputtered on \ch{SiO_2} (sample $O_5$) and on SiN (sample $N_1$); both samples were sputtered simultaneously. (c) Rocking curves around the AlN~(0002) reflection for AlN thin films on SiN with different thicknesses deposited. The curves are normalized using their fitted peak height. The extracted FWHM are plotted against the film thickness in the inset. Rocking curves are shifted to \SI{0}{\degree}.}
\end{figure}

The sputtering conditions of sample $O_4$ were also used to grow AlN on SiN. A comparative experiment is performed by a sputter deposition run on \ch{SiO_2/Si} (sample $O_5$) and \ch{Si_3N_4/SiO_2/Si} (sample $N_1$). The $2\theta$-$\omega$ scan and the rocking curve of the AlN~(0002) reflection are shown in Figure \ref{fig:compare_SiN_and_SiO2}(a) and Figure \ref{fig:compare_SiN_and_SiO2}(b), respectively. The $2\theta$-$\omega$ scans show comparable intensities of the (0002) reflections indicating similar crystal properties between $N_1$ and $O_5$. The rocking curve also shows a similar FWHM of both thin films, thus indicating an equivalent mosaic spread of the c-axis orientation.

For integration into photonic circuits, we are interested in studying the quality of ultra-thin AlN films. The SiN waveguides in our photonic circuits have a thickness of about \SI{330}{nm} \cite{poot2016characterization}, therefore similar or thinner thicknesses are used to realize hybrid AlN/SiN waveguides. 
In particular, we estimate that for AlN films thicker than \SI{110}{nm}, slab modes \cite{saleh_teich} can propagate through the thin film, whereas for thinner films these are not supported. Thus, the film thickness can be used to tune the propagation properties in hybrid SiN/AlN waveguides. We compared four different AlN thicknesses ranging from \SI{297}{nm} to \SI{74}{nm} (samples $N_1$ to $N_4$). $2\theta$-$\omega$ scans indicate that all AlN films have again a strong c-axis orientation (not shown). Next, the rocking curves around the AlN~(0002) reflections are measured and fitted using a Gaussian with a constant background (see Figure~\ref{fig:compare_SiN_and_SiO2}(c)). The obtained FWHMs are listed in Table \ref{tab:table1deposition}. Thinner films have a lower intensity due to the smaller film thickness, and have a larger FWHM. The FWHM values are plotted against thickness in the inset of Figure \ref{fig:compare_SiN_and_SiO2}(c). They show a trend that is inversely proportional to the thickness, represented by the black line in the inset. Similar thickness dependencies of the FWHM have already been observed for AlN on SiN \cite{Belkerk2014} and on molybdenum \cite{Felmetsger2011}. Therefore, a larger FHWM is expected for thinner films. However, these AlN thin films are still c-axis oriented as also revealed by $2\theta$-$\omega$ scans. 

\subsection{Surface properties}
\begin{figure}[ht]
\centering
\includegraphics[width=1\textwidth]{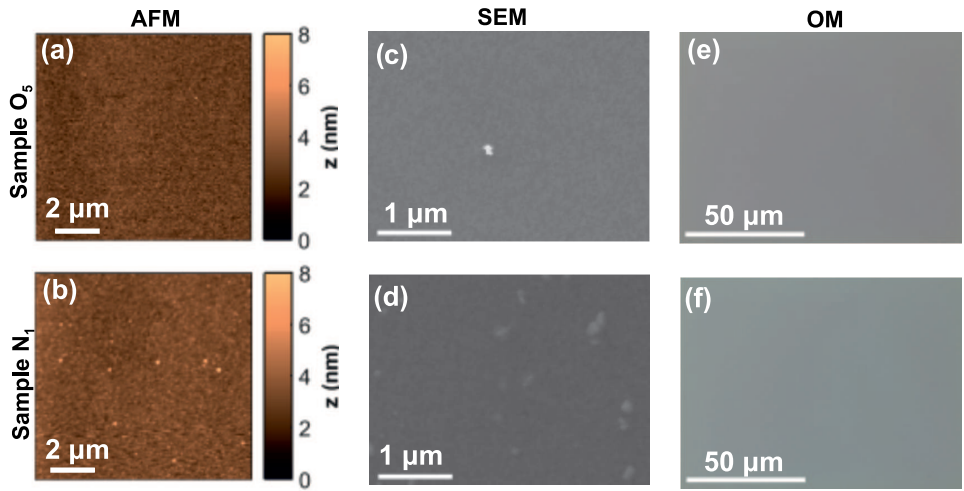}
\caption{\label{fig:films}(a-b) $10\times$\SI{10}{\mathbf{\mu}m^2} AFM scans with 256 points per line of samples $O_5$ and $N_1$. Extracted RMS roughnesses are \SI{0.64}{nm} and \SI{0.47}{nm}, respectively. SEM images (c-d) and optical micrographs (e-f) of the same two samples. Gold particles are placed on the surface to have a focusing point in (c-d). The films do not show any features, thus confirming smooth surface and homogeneous optical properties.}
\end{figure}

An important aspect of the deposited films, especially for photonic applications, are their surface properties. The RMS surface roughness of the substrates and of the deposited films is characterized using a \textit{Nanosurf CoreAFM}. A $10 \times \SI{10}{\mathbf{\mu}m^2}$ scan is performed at the center of the film, and the RMS surface roughness is extracted using the software \textit{Gwyddion}. The AFM scans for samples $O_5$ and $N_1$ are shown in Figure \ref{fig:films}(a) and (b), respectively. The AlN thin film sputtered on \ch{SiO_2} shows a RMS roughness of \SI{0.64}{nm}, while the one on SiN shows a RMS roughness of \SI{0.47}{nm}. The obtained values for the other samples are listed in Table~\ref{tab:table1deposition}. All fabricated films have a RMS roughness below \SI{1}{nm}, which is excellent for photonic application due to decreased light scattering losses. Furthermore, the RMS roughness is below common literature values, which are often above \SI{1}{nm}\cite{Felmetsger2011,Larciprete2006,Iqbal2016}. 
The AlN surface is also characterized via SEM. Two representative SEM images are shown in Figure~\ref{fig:films}(c) and \ref{fig:films}(d) of samples $O_5$ and $N_1$, respectively. Gold particles are placed on the surface to help focusing and are the only visible features on the scan. Therefore, the SEM images confirm the low surface roughness. Moreover, samples $O_5$ and $N_1$ are also characterized using an optical microscope and their optical micrographs are displayed in Figure \ref{fig:films}(e) and (f), respectively. The films do not have any visible features, thus showing homogeneous optical properties.  

\subsection{Spectral reflectance}
\begin{figure}[ht]
\centering
\includegraphics[width=0.58\textwidth]{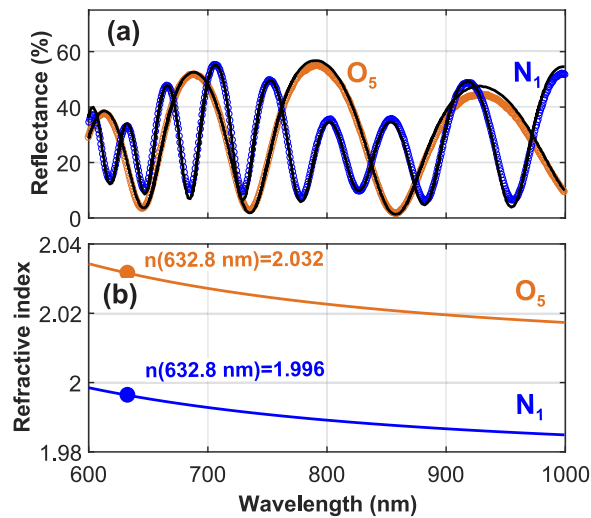}
\caption{\label{fig:reflectometry}(a) Reflectometry of samples $O_5$ and $N_1$. The experimental points are fitted by the \textit{Filmetrics F40} thin-film analyzer (black solid line). (b) Extracted refractive indices, with constrained Cauchy model dependence against wavelength.}
\end{figure}

The optical properties of the sputtered AlN are further characterized via spectral reflectance. Spectral reflectance, also known as reflectometry, is an optical technique for thin-film characterization: the light reflected from a thin film is measured over a range of wavelengths, with the incident light perpendicular to the surface. The reflectance spectrum depends on the optical constants and the thickness of the film stack. These properties can be derived by fitting the measured spectrum. Figure \ref{fig:reflectometry}(a) shows the measured reflection of samples $O_5$ and $N_1$ with their respective fits. The spectra are measured and are fit using a \textit{Filmetrics F40} thin-film analyzer. The oscillations are due to constructive and destructive interference from the reflections at the different interfaces of the films (air-AlN, AlN-SiN, SiN-\ch{SiO_2}, etc.). The fitting software is provided with rough starting values for the AlN thickness $h_\text{AlN}$, for the AlN refractive index $n_\text{AlN}$ and for the extinction coefficient $k_\text{AlN}$. The substrate layer stack is known ($O_5$ and $N_1$ have \SI{1500}{nm} \ch{SiO_2} on Si, and \SI{330}{nm} SiN, \SI{3300}{nm} \ch{SiO_2} on Si, respectively). The refractive index $n_\text{AlN}(\lambda)$ is set to follow a Cauchy model. The fitted $n_\text{AlN}(\lambda)$ are shown in Figure~\ref{fig:reflectometry}(b), while the obtained indices at \SI{632.8}{nm} for all samples are listed in Table \ref{tab:table1deposition}. The values are in the $n=2.00-2.09$ interval, which is close to the expected value for sputtered highly c-axis oriented AlN \cite{Ababneh2020,Larciprete2006}, that ranges between $n=2.0-2.2$ depending on internal stress \cite{xiong2012aluminum}. It can be noted that the extracted refractive indices of the films grown on SiN are usually lower than the ones for \ch{SiO_2}. Therefore, the two substrates might have a different influence on the film stress. The difference in refractive index can also be observed from the different film color in Figure ~\ref{fig:films}(e-f). The absorption coefficient $k_\text{AlN}(\lambda)$ is also obtained. However, the value is found to be close to zero and below the resolution of the thin-film analyzer. This indicates that, at least in the visible and near infrared (NIR) wavelength ranges, the deposited AlN has a low absorption. Furthermore, the refractive index is similar to \ch{Si_3N_4} (2.1). This similarity, combined with the low absorption, makes AlN promising for our hybrid integration approach where AlN will be grown on top of prefabricated SiN photonic circuits \cite{Terrasanta2021} for a wide range of application in photonics, optomechanics, and integrated quantum optics.

\subsection{Hybrid photonic circuit}
\begin{figure}[ht]
\centering
\includegraphics[width=0.95\textwidth]{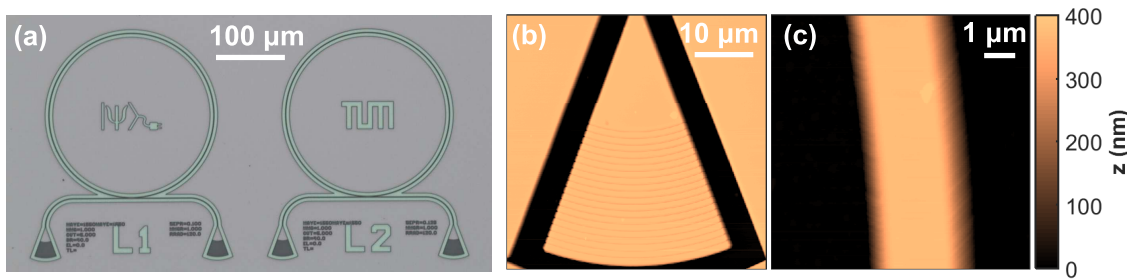}
\caption{\label{fig:hybrid}(a) An optical micrograph of two hybrid microring resonators with an AlN thickness of \SI{204}{nm}. AFM scans of a grating coupler (b) and of a waveguide (c) in a hybrid photonic circuit with an AlN thickness of \SI{204}{nm}. 
The extracted RMS surface roughness is \SI{0.26}{nm} and \SI{0.51}{nm} on protected (copper-colored) and etched regions (black), respectively.}
\end{figure}

The starting point of the aforementioned hybrid photonic circuit fabrication is the patterning of SiN. First, \textit{ZEP520A} resist is spin-coated and photonic circuitry is defined using an electron beam lithography system (\textit{Nanobeam nB5}). The pattern is transferred into the SiN by reactive ion etching in an inductively-coupled plasma etcher (\textit{Oxford PlasmaPro 80 ICP}) using an \ch{SF_6/CHF_3} chemistry. The pattern is etched all the way into the cladding layer, with an etch depth in the \ch{SiO_2} of about \SI{50}{nm}. The remaining resist is stripped and as the final step, AlN is sputtered on top of the patterned SiN with the same sputtering conditions as for the $N_i$ samples. An optical micrograph of hybrid microring resonators made using this process is shown Figure \ref{fig:hybrid}(a). In addition, hybrid structures with \SI{204}{nm} of AlN are characterized by AFM. The scans of a grating coupler, which can be used to couple light from a fiber to the in-plane waveguide \cite{Taillaert2002}, and of a waveguide are shown in Figure \ref{fig:hybrid} (b) and (c), respectively. The RMS surface roughness is extracted from the AlN deposited on the etched regions, namely on \ch{SiO_2}. The obtained value is \SI{0.51}{nm}, which is similar to the values for AlN deposited on the pristine \ch{SiO_2} of samples $O_i$; see Table \ref{tab:table1deposition}. The surface roughness is thus not affected by the fabrication steps applied before the AlN sputtering, resulting in high-quality hybrid photonic structures \cite{Terrasanta2021}.

\section{Conclusion and outlook}
AlN c-axis thin films were successfully fabricated by DC reactive magnetron sputtering. First, the growth on \ch{SiO_2} was improved by optimizing the N$_2$/Ar flow ratio and the sputtering power. Structural characterization by XRD shows that (0001)-oriented films with an AlN (0002) rocking curve FWHM of \SI{3.70}{\degree} at a film thickness of \SI{238}{nm} can be achieved after optimization. Afterwards, the optimized conditions were employed for the sputter deposition of AlN on top of SiN with different thicknesses. The XRD measurements confirm the dominant c-axis orientation, with a rocking curve FWHM ranging from \SI{3.93}{\degree} at \SI{297}{nm} to \SI{6.81}{\degree} at \SI{74}{nm}, and thus proving the possibility of fabricating highly-crystalline ultrathin AlN films on top of SiN. The surface roughness of the films was analyzed by AFM. For the growth on both SiN and \ch{SiO_2} films small RMS roughness values below \SI{1}{nm} have been demonstrated. The smooth surface was also confirmed by SEM. Finally, the optical properties were characterized by spectral reflectance. The obtained refractive indices are in the expected range, while the extinction coefficient is below the detection limit. \\
We have employed this fabrication to realize hybrid photonic circuits. To do so, AlN was deposited on top of chips that had  pre-patterned SiN photonic circuitry. For instance, we have realized AlN/SiN composite microring resonators, which showed improved optical properties after the deposition, such as lower bending losses and lower propagation losses \cite{Terrasanta2021}. As the integrated films are c-axis oriented, this novel composite approach will be a promising solution for the integration of second-order nonlinear optical properties in SiN photonic circuits. The next step would be to find the AlN thickness that is needed to realize phase-matched waveguides and test the nonlinear processes, such as second-harmonic generation, and the generation of single photon pairs using spontaneous parametric down conversion. Moreover, a systematic deposition temperature optimization could be performed to achieve further improvements of the crystal orientation at the chosen thickness.

\ack
We thank David Hoch and Xiong Yao for their help with the photonic circuitry nanofabrication. This project is funded by the German Research Foundation (DFG) under Germany's Excellence Strategy - EXC-2111 - 390814868 and TUM-IAS, funded by the German Excellence Initiative and the European Union Seventh Framework Programme under grant agreement 291763.

\section*{References}
\bibliography{AlNsputtering}

\providecommand{\newblock}{}
\begin{thebibliography}{10}
\expandafter\ifx\csname url\endcsname\relax
  \def\url#1{{\tt #1}}\fi
\expandafter\ifx\csname urlprefix\endcsname\relax\def\urlprefix{URL }\fi
\providecommand{\eprint}[2][]{\url{#2}}

\bibitem{trolier2004thin}
Trolier-McKinstry S and Muralt P 2004 {\em Journal of Electroceramics\/} {\bf
  12} 7--17 \urlprefix\url{https://doi.org/10.1023/B:JECR.0000033998.72845.51}

\bibitem{Piazza2006}
{Piazza} G, {Stephanou} P~J and {Pisano} A~P 2006 {\em Journal of
  Microelectromechanical Systems\/} {\bf 15} 1406--1418

\bibitem{Dubois1999}
Dubois M~A and Muralt P 1999 {\em Applied Physics Letters\/} {\bf 74}
  3032--3034 \urlprefix\url{https://doi.org/10.1063/1.124055}

\bibitem{wan2020large}
Wan N~H, Lu T~J, Chen K~C, Walsh M~P, Trusheim M~E, De~Santis L, Bersin E~A,
  Harris I~B, Mouradian S~L, Christen I~R {\em et~al.\/} 2020 {\em Nature\/}
  {\bf 583} 226--231 \urlprefix\url{https://doi.org/10.1038/s41586-020-2441-3}

\bibitem{xiong2012aluminum}
Xiong C, Pernice W~H~P, Sun X, Schuck C, Fong K~Y and Tang H~X 2012 {\em New
  Journal of Physics\/} {\bf 14} 095014
  \urlprefix\url{https://doi.org/10.1088/1367-2630/14/9/095014}

\bibitem{pernice2012}
Pernice W~H~P, Xiong C, Schuck C and Tang H~X 2012 {\em Applied Physics
  Letters\/} {\bf 100} 223501 \urlprefix\url{https://doi.org/10.1063/1.4722941}

\bibitem{guo2017parametric}
Guo X, Zou C~l, Schuck C, Jung H, Cheng R and Tang H~X 2017 {\em Light: Science
  \& Applications\/} {\bf 6} e16249--e16249
  \urlprefix\url{https://doi.org/10.1038/lsa.2016.249}

\bibitem{Liu_OpEx_AlN_on_sapphire_UV}
Liu X, Bruch A~W, Gong Z, Lu J, Surya J~B, Zhang L, Wang J, Yan J and Tang H~X
  2018 {\em Optica\/} {\bf 5} 1279--1282
  \urlprefix\url{http://www.osapublishing.org/optica/abstract.cfm?URI=optica-5-10-1279}

\bibitem{Jung:14}
Jung H, Stoll R, Guo X, Fischer D and Tang H~X 2014 {\em Optica\/} {\bf 1}
  396--399
  \urlprefix\url{http://www.osapublishing.org/optica/abstract.cfm?URI=optica-1-6-396}

\bibitem{Larciprete2006}
Larciprete M~C, Bosco A, Belardini A, {Li Voti} R, Leahu G, Sibilia C, Fazio E,
  Ostuni R, Bertolotti M, Passaseo A, Pot{\`{i}} B and {Del Prete} Z 2006 {\em
  Journal of Applied Physics\/} {\bf 100} 023507

\bibitem{Dasgupta2009}
Dasgupta S, Wu F, Speck J~S and Mishra U~K 2009 {\em Applied Physics Letters\/}
  {\bf 94} 151906 \urlprefix\url{https://doi.org/10.1063/1.3118593}

\bibitem{chiang2011}
Chiang C, Chen K, Wu Y, Yeh Y, Lee W, Chen J, Lin K, Hsiao Y, Huang W and Chang
  E 2011 {\em Applied Surface Science\/} {\bf 257} 2415 -- 2418 ISSN 0169-4332
  \urlprefix\url{http://www.sciencedirect.com/science/article/pii/S016943321001425X}

\bibitem{szekeres2011}
Szekeres A, Fogarassy Z, Petrik P, Vlaikova E, Cziraki A, Socol G, Ristoscu C,
  Grigorescu S and Mihailescu I 2011 {\em Applied Surface Science\/} {\bf 257}
  5370 -- 5374 ISSN 0169-4332
  \urlprefix\url{http://www.sciencedirect.com/science/article/pii/S0169433210014091}

\bibitem{Iqbal2018}
Iqbal A and Mohd-Yasin F 2018 {\em Sensors\/} {\bf 18} 1797
  \urlprefix\url{https://doi.org/10.3390/s18061797}

\bibitem{hoch2020chip}
Hoch D, Sommer T, Mueller S and Poot M 2020 {\em Turkish Journal of Physics\/}
  {\bf 44} 239--246

\bibitem{poot2016design}
Poot M, Schuck C, song Ma X, Guo X and Tang H~X 2016 {\em Opt. Express\/} {\bf
  24} 6843--6860
  \urlprefix\url{http://www.opticsexpress.org/abstract.cfm?URI=oe-24-7-6843}

\bibitem{poot2016characterization}
Poot M and Tang H~X 2016 {\em Applied Physics Letters\/} {\bf 109} 131106
  \urlprefix\url{https://doi.org/10.1063/1.4962902}

\bibitem{Terrasanta2021}
Terrasanta G, Sommer T, M{\"u}ller M, Althammer M and Poot M 2021 {Aluminum
  nitride integration on silicon nitride photonic circuits: a new hybrid
  approach towards on-chip nonlinear optics} (In Preparation)

\bibitem{depla2019}
Depla D, Strijckmans K, Dulmaa A, Cougnon F, Dedoncker R, Schelfhout R, Schramm
  I, Moens F and {De Gryse} R 2019 {\em Thin Solid Films\/} {\bf 688} 137326
  ISSN 0040-6090 a Special Issue “Thin Film Advances”, dedicated to the
  75th birthday of Professor Joe Greene
  \urlprefix\url{http://www.sciencedirect.com/science/article/pii/S0040609019303232}

\bibitem{Iqbal2016}
Iqbal A, Walker G, Iacopi A and Mohd-Yasin F 2016 {\em Journal of Crystal
  Growth\/} {\bf 440} 76 -- 80 ISSN 0022-0248
  \urlprefix\url{http://www.sciencedirect.com/science/article/pii/S0022024816300197}

\bibitem{levinshtein2001properties}
Levinshtein M~E, Rumyantsev S~L and Shur M~S 2001 {\em Properties of Advanced
  Semiconductor Materials: GaN, AIN, InN, BN, SiC, SiGe\/} (John Wiley \& Sons)

\bibitem{Downs2003}
Downs R and Hall-Wallace M 2003 {\em American Mineralogist\/} {\bf 88} 247--250
  ISSN 16113349

\bibitem{Xu2001}
Xu X~H, Wu H~S, Zhang C~J and Jin Z~H 2001 {\em Thin Solid Films\/} {\bf 388}
  62 -- 67 ISSN 0040-6090
  \urlprefix\url{http://www.sciencedirect.com/science/article/pii/S0040609000019143}

\bibitem{xrd}
Als-Nielsen J and McMorrow D 2011 {\em {Elements of Modern X-ray Physics}\/}
  ({John Wiley \& Sons}) ISBN 978-1-119-99836-5

\bibitem{Felmetsger2011}
Felmetsger V~V, Laptev P~N and Graham R~J 2011 {\em Journal of Vacuum Science
  \& Technology A\/} {\bf 29} 021014
  \urlprefix\url{https://doi.org/10.1116/1.3554718}

\bibitem{saleh_teich}
Saleh B~A~E and Teich M~C 1991 {\em Fundamentals of Photonics\/} (John Wiley
  and Sons)

\bibitem{Belkerk2014}
Belkerk B~E, Bensalem S, Soussou A, Carette M, Al~Brithen H, Djouadi M~A and
  Scudeller Y 2014 {\em Applied Physics Letters\/} {\bf 105} 221905
  \urlprefix\url{https://doi.org/10.1063/1.4903220}

\bibitem{Ababneh2020}
Ababneh A, Albataineh Z, Dagamseh A, Al-kofahi I, Schäfer B, Zengerle T, Bauer
  K and Seidel H 2020 {\em Thin Solid Films\/} {\bf 693} 137701 ISSN 0040-6090
  \urlprefix\url{http://www.sciencedirect.com/science/article/pii/S004060901930728X}

\bibitem{Taillaert2002}
{Taillaert} D, {Bogaerts} W, {Bienstman} P, {Krauss} T~F, {Van Daele} P,
  {Moerman} I, {Verstuyft} S, {De Mesel} K and {Baets} R 2002 {\em IEEE Journal
  of Quantum Electronics\/} {\bf 38} 949--955

\end{thebibliography}
\end{document}